\documentclass[twocolumn,showpacs,preprintnumbers,amsmath,amssymb]{revtex4}

\usepackage{graphicx}
\usepackage{dcolumn}
\usepackage{bm}

\begin{document}

\title{Magnetic domain walls displacement : automotion vs. 
spin-transfer torque}
\author{Jean-Yves Chauleau}
\author{Rapha{\"{e}}l Weil}
\author{Andr{\'{e}} Thiaville}%
\author{Jacques Miltat}
\affiliation{Laboratoire de Physique des Solides, CNRS UMR 8502, 
Univ. Paris-sud, 91405 Orsay Cedex, France}%



\begin{abstract}
The magnetization dynamics equation predicts that a domain
wall that changes structure should undergo a displacement by
itself - automotion - due to the relaxation of the linear 
momentum that is associated with the wall structure.
We experimentally demonstrate this effect in soft nanostrips, 
transforming under spin transfer torque a metastable asymmetric 
transverse wall into a vortex wall.
Displacements more than three times as large as under spin 
transfer torque only are measured for 1~ns pulses.
The results are explained by analytical and numerical micromagnetics.
Their relevance to domain wall motion under spin transfer torque is 
emphasized.

\end{abstract}


\maketitle
The displacement of magnetic domain walls (DW) by
spin transfer torque (STT \cite{Berger84}) 
is presently considered 
as a means of control in device applications \cite{Parkin08}.
It also raises fundamental questions about the description
of electronic transport in magnetic media
\cite{Tserkovnyak08}, fostering many experimental 
\cite{Yamaguchi04,Klaui05,Meier07,Vlaminck08}
and theoretical
\cite{Bazaliy98,Zhang04,Tatara04}
studies.
The former can be divided into two groups, according to the duration
of current application.
With long ($\approx \mu$s) pulses, average wall
velocities much lower than micromagnetic expectations 
\cite{Thiaville05} have been first observed \cite{Yamaguchi04,Jubert06}.
In this regime, sample heating limits the applicable current
densities, and wall motion occurs under a strong influence of pinning
\cite{Yang07}.
For short ($\approx$~ns) pulses however, higher apparent
DW velocities have been reported \cite{Hayashi07,Meier07,Heyne10}, 
interpreted by an easier depinning due to an additional force
on the DW during the pulse risetime \cite{Bocklage09}.
The measurement of the DW \emph{velocity} under STT is of importance
because theoretical predictions relate the fundamental parameters
of spin-polarized electron transport to the initial \cite{Li04,Thiaville07}
and steady-state \cite{Zhang04,Thiaville05} DW velocities.

Under pulsed field, however, another type of DW motion has been 
documented, known as streaming or gyromagnetic \cite{Stein67}, 
overshoot or automotion \cite{Malozemoff79}, depending on the 
field direction.
The common ingredient to these situations is that certain changes
of the wall structure lead to a wall displacement.
This is ultimately related to the fact \cite{Thiele76,Slonczewski79}
that a characteristics of the wall structure (equivalent to an 
angle of the wall magnetization)
plays the role of a linear momentum conjugated, in Hamilton's sense,
to the wall position.
Even if these earlier works considered Bloch walls, where the concept
of a wall angle is intuitive, it was recently shown to apply to any type of
wall \cite{Thiaville07}.
As changes of the wall structure after a current pulse have been
observed in some cases \cite{Klaui05,Parkin08,Heyne08b}, it is important 
to experimentally evaluate the wall displacement purely due to this change.
In particular, does automotion persist in presence of the unavoidable 
sample imperfections that pin the wall, and what is the signature
of this phenomenon ?
This is the object of the experiments described in this Letter.
We apply high resolution magnetic force microscopy (MFM) to
observe wall structure and position in permalloy nanostrips, before and 
after sharp current pulses of nanosecond duration.
Automotion is demonstrated, with a large DW displacement in a direction 
related to the sense of the wall angle change, rather than to current polarity.
The observations are interpreted, qualitatively and quantitatively, 
by micromagnetics. 
\begin{figure}[b]
\scalebox{0.4}{\includegraphics{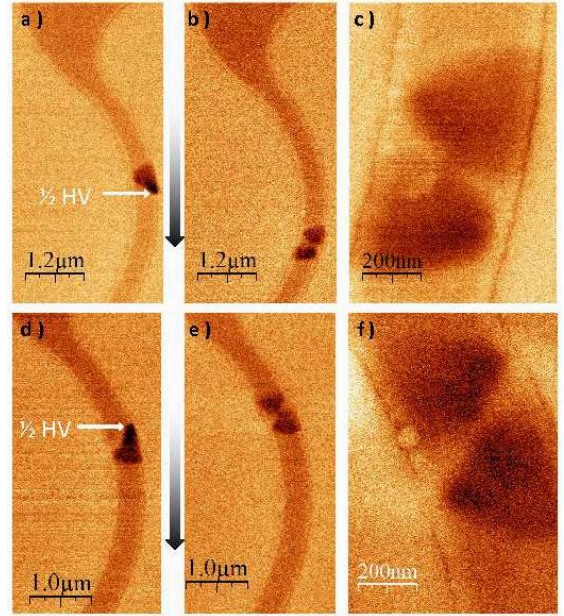}}
\caption{(color online) Automotion demonstrated by MFM imaging. 
An ATW (a) with positive asymmetry turns into a VW (b) 
after one current pulse (1~ns, 2.4~TA/m$^{2}$)
with a displacement of 1.4~$\mu$m in the same direction as the electron flow
(shaded arrow). 
The zoom on the core of this VW (c) shows the continuity of the dark contrast. 
Another ATW with negative asymmetry (d) turns into a VW (e) after one 
same pulse, with a displacement of 
0.7~$\mu$m, now in the direction opposite to electron flow. 
The zoom on the vortex core (f) indicates that the bright contrast is
continuous.
The half hedgehog vortices (1/2~HV) are indicated in (a) and (d). }
\label{fig:automotion}
\end{figure}
%

Observations were performed in a MFM fitted with RF connection in order 
to apply short current pulses (rise and falltimes below 100~ps).
Special care was taken for decreasing and controlling the 
magnetic perturbation due to the tip. 
First, the silicon tips were coated with a very thin 
Co$_{70}$Cr$_{30}$ 4.5~nm layer with low coercivity, so that the
tip reverses under the DW stray field. 
This means that the tip-sample interaction is always attractive, as 
testified by dark DW MFM contrasts. 
Second, two actions of the tip were sometimes observed, in which the DW 
was either snatched at tip approach, or dragged along with the tip. 
The snatch-up case, occuring for a tip $<100$~nm away from
the DW position (consistent with the 10~Oe DW propagation field 
measured for this sample), little affects the scatter of the measured
DW displacements.
In the cases of dragging (10\% frequency), the displacement incurred 
was subtracted from the raw displacement. 

Samples were prepared from a magnetically soft 
Pd(3.5)/Ni$_{80}$Fe$_{20}$(17.5)/Pd(3.5) layer (thicknesses in nanometers) 
patterned by e-beam lithography and lift-off 
into nanostrips $w = 450$~nm wide and 12~$\mu$m long. 
A Ti(3)/Au(100) coplanar waveguide connecting the sample 
was fabricated in a second step.
Given the samples width and thickness, vortex walls (VW)  are 
energetically stable \cite{Nakatani05}. 
Nevertheless, the DW initial state prepared by saturating the sample with a 
strong transverse field ($\approx$1~kOe) is a metastable \cite{Klaui04}
asymmetric transverse wall (ATW, Fig.~\ref{fig:automotion}a,d). 
Once the created ATW is imaged, a current pulse, 1~ns long and in the 
range of amplitudes for STT (a few TA/m$^2$), is applied. 
The transformed DW is then imaged, revealing its detailed structure and
displacement.
Samples have the shape of an `S' so that two different walls are
simultaneously nucleated.

Fig.~\ref{fig:automotion} shows typical results, where the ATW transformed
to the stable VW structure.
In every case, a DW displacement was observed,
either along the direction of electron flow (a-b) or in the opposite
direction (d-e).
The typical displacement, $\approx$1~$\mu$m, would correspond to a
1~km/s effective velocity, a very large value considering the current
density of 2.4~TA/m$^2$ (equivalent to a 84~m/s spin drift 
velocity \cite{Thiaville05}).
Moreover, displacement in both directions is observed, which is not 
consistent with STT.
The direction of displacement is however not random.
For example, the two ATW in Fig.~\ref{fig:automotion}, created by
the same transverse field, have the same domain and wall magnetization 
directions, but opposite asymmetries.
The asymmetry, seen as an inclination of the contrast along the nanostrip 
length, in either direction, appears randomly upon nucleation.
We also remark that the polarities of the vortex cores in the final states
are opposite (this polarity appears in the MFM images, as a continuity
or an interruption of the contrast of the two wings of the VW, due to the
superposition of DW and vortex core magnetic charges).

\begin{figure}[b]
\includegraphics[width=0.7\columnwidth]{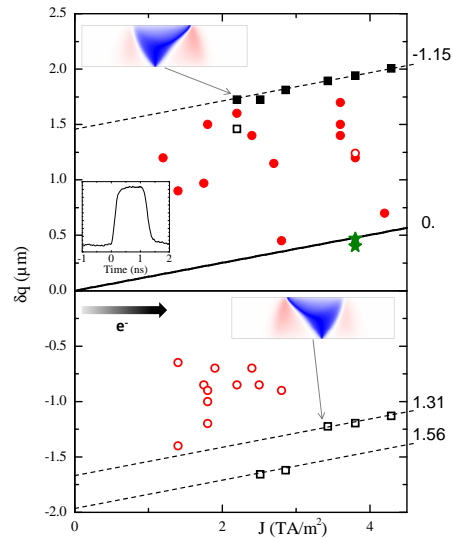}
\caption{(color online)
Compilation of DW displacements $\delta q$ measured after
ATW to VW transformations on different samples (pulse shape shown
in inset).  
The electron flow defines the positive axis along the nanostrip.
Filled (open) circles denote positive (resp. negative) asymmetry
of the initial ATW, see the inset images (colored
according to the transverse magnetization component).
Squares show the micromagnetic simulation results.
The solid line shows the computed pure STT displacement, 
to which some experimental data (stars) for a VW, without transformation,
have been superposed.
Dashed lines link points with similar calculated $\delta \Phi$.
}
\label{fig:gene}
\end{figure}

We study these phenomena with the Landau-Lifschitz-Gilbert
equation for magnetization dynamics, augmented with the STT terms
\cite{Zhang04,Thiaville05}. 
The analytical analysis capturing the physics of the DW displacement
upon transformation is first recalled.
As the relaxation time of the vortex wall structure is much longer than 
the current pulse duration, the transformation essentially takes place 
without current.
The dynamics of a DW structure in the absence of any external torque
is termed automotion.
For the geometry of a DW in a nanostrip, it has been shown that, whatever 
the wall type, the following equation then holds \cite{Thiaville07}:
\begin{equation}
\label{eq:1Deff}
\frac{d \Phi}{d t} + \frac{\alpha}{\Delta_\mathrm{T}}
\frac{d q}{d t} = 0,
\end{equation}
where $\Phi$ is the generalized wall magnetization angle, $q$
the (generalized) wall position, and $\Delta_\mathrm{T}$ the
Thiele domain wall width.
When the DW structure is planar and transforms by the motion
of one vortex, the DW displacement obtained by time
integration )under the assumption of a constant DW width) reads  
$\delta q =  p \delta y_\mathrm{c} \pi \Delta_\mathrm{T}/(\alpha w)$. 
Its sign is fixed by the vortex core polarity $p$ and 
by the path followed by the core (the change of its position 
$y_\mathrm{c}$ across the width $w$ of the nanostrip).
The three quantities $\delta q$, $p$ and $\delta y_\mathrm{c}$ are 
directly observed by MFM.
For the last quantity, one should notice that the ATW structure has a 
position where a vortex is most easily injected, namely the
half hedgehog vortex (1/2~HV, see Ref.~\cite{Thiaville09}), indicated
in Fig.~\ref{fig:automotion}.
A proof of this injection path is obtained by comparing the inclination
of the stripe in-between the two wings of the VW to the inclination
of the ATW.
As the initial $y_\mathrm{c}$ is 0 or $w$ and the final $w/2$,
$\delta y_\mathrm{c}$ is $\pm w/2$.
It is observed that the signs of all displacements observed under ATW 
to VW transformations are explained by this simple relation.

\begin{figure}[b]
\includegraphics[width=\columnwidth]{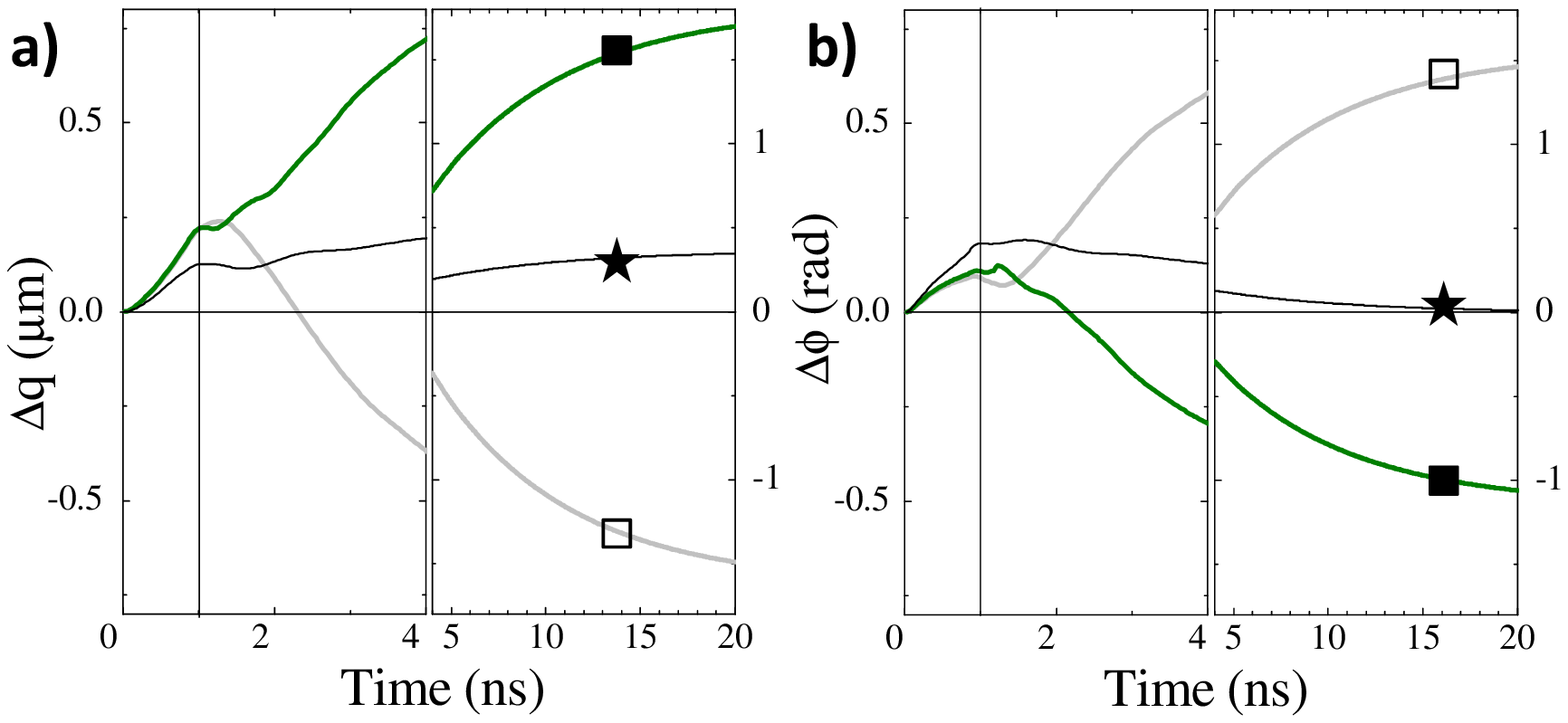}
\caption{(color online) Numerical micromagnetic calculation, for
positive (filled square) and negative asymmetry ATW (open square) as 
well as VW (star), of the effect of a 1~ns, 2.86~TA/m$^2$ current pulse.
The wall displacement (a) and generalized wall angle change (b) are 
plotted.
 }
\label{fig:calc1}
\end{figure}

The magnitude of the displacements is now discussed.
The measurements, on 8 different samples with slightly 
varying widths $w= 450-500$~nm, for different current directions
and ATW asymmetries, are gathered in Fig.~\ref{fig:gene}.
Both positive (along the electron flow) and negative DW
displacements are observed, the sign correlating 
with the ATW asymmetry, except in one case.
The displacements are very different from those 
expected for a pure STT effect (shown by the solid line).
On the other hand, 
the calculated Thiele DW width being $\Delta_\mathrm{T}= 26$~nm
for a $450 \times 17.5$~nm$^2$ nanostrip \cite{Thiaville07}, the
analytical model predicts a displacement 
$(\pi / 2) \Delta_\mathrm{T} / \alpha= 2.04$~$\mu$m for
a damping $\alpha=0.02$.
Numerical micromagnetic computations were also performed, using a 
homemade code \cite{Miltat07} adapted to
the infinite nanostrip geometry with a moving calculation box 
centered on the DW.
Parameters were: current polarization $P= 0.5$, non-adiabatic 
STT coefficient $\beta= 0.08$, 
damping constant $\alpha= 0.02$, and mesh size 
$3.68 \times 3.68 \times 17.5$~nm$^3$.
The Oersted field created by the current was included.
These numerically computed displacements are reported in 
Fig.~\ref{fig:gene}.
The computed minimum current density for DW transformation
is 2.3~TA/m$^2$, a value above the experimental result, however
obtained for a perfect nanostrip (it was even larger for lower $\beta$, 
or when neglecting the Oersted field). 
The calculations also show the clear correlation of displacement
sign with ATW asymmetry.
Regarding magnitudes, one observes that the largest 
experimental DW displacements are close to calculations, using
the effective value $\alpha= 0.02$ that has been recognized as
appropriate for DW dynamics in NiFe \cite{Nakatani03,Min10}.
Smaller experimental displacements are ascribed 
to sample imperfections pinning the DW.

In addition, the calculations reveal that displacements increase 
with current density, with a slope similar to the STT contribution. 
However, the data form groups with different zero-current 
extrapolated displacements, only the largest reaching
the analytical value.
This shows that the ATW transformation is in fact complex.
Indeed, calculations show that vortex cores sometimes reverse
polarity, this mostly happening shortly after vortex injection,
when the driving force towards the nanostrip center causes
large vortex core velocities.
In such cases, the $\delta q$ contributions of 
all cores have to be added (with a minus sign for antivortices), 
leading to reduced and sometimes reversed displacements as
well as to reduced values of $\delta \Phi$ 
(see Fig.~\ref{fig:gene}).

\begin{figure}[b]
\includegraphics[width=\columnwidth]{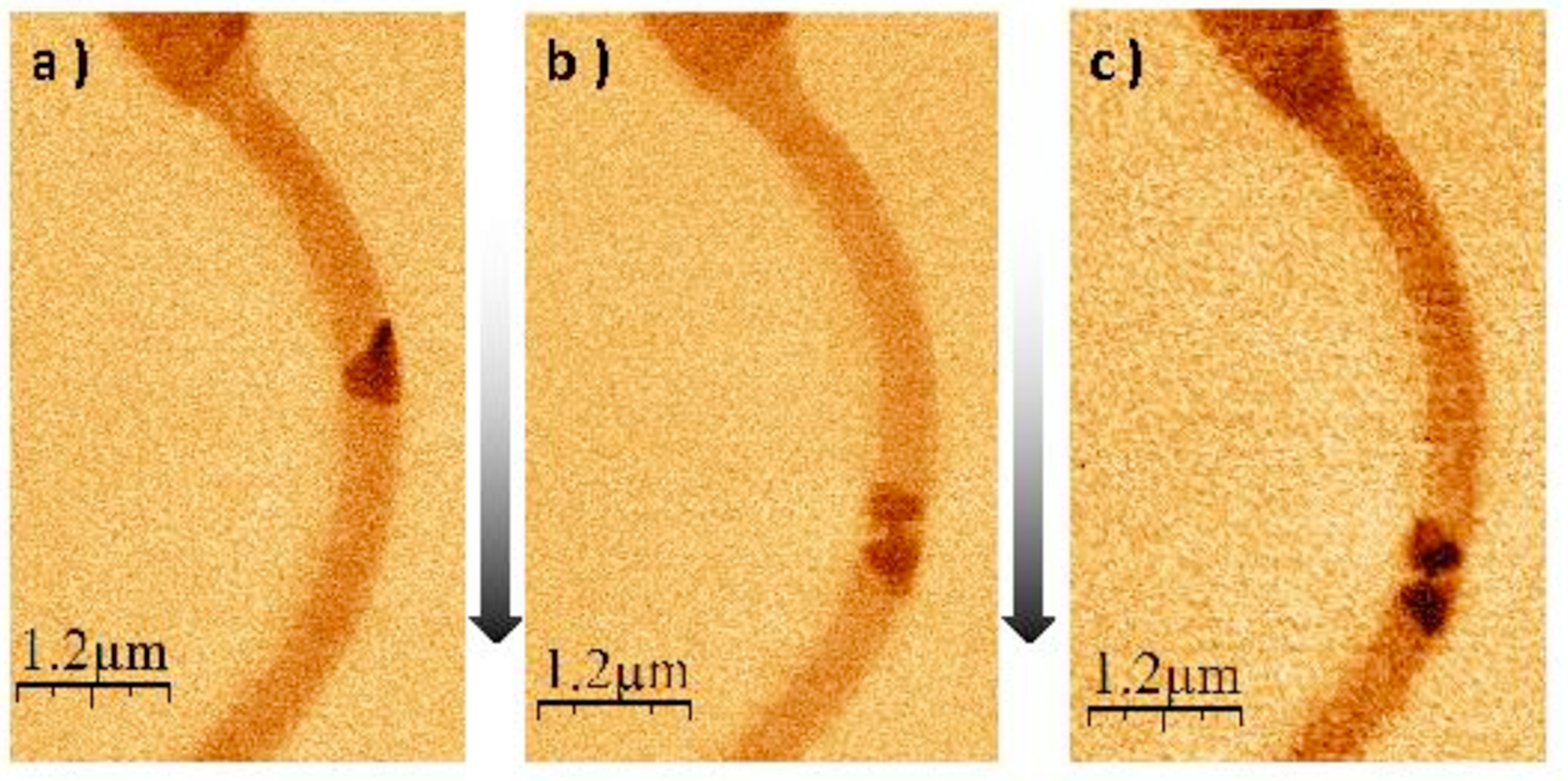}
\caption{(color online) Comparison of automotion and STT induced DW 
displacements. 
An ATW (a) turns into a VW (b) after one pulse of (1~ns, 3.8~TA/m$^2$) with a 
displacement of 1.25~$\mu$m in the same direction as the STT action. 
This VW displaces under STT by 400~nm (c) with a
second pulse.
 }
\label{fig:stt}
\end{figure}

Fig.~\ref{fig:calc1} details the numerical results 
for the situation of Fig.~\ref{fig:automotion}.
The ATWs with opposite asymmetries finally move
in opposite directions, and magnetization snapshots show that 
vortices of opposite polarities finally appear, even if
identical vortices are initially injected at the 1/2~HV positions
\cite{online-movies}.
Note that the displacements extend over 20~ns, which corresponds to
the large relaxation time of the VW (7.5~ns according to 
Ref.~\cite{Thiaville07}).
In addition, the figure also shows the results for a VW: 
the angle $\Phi$ increases during the pulse (less than for the
ATW as the DW relaxation time is larger), and then goes back to zero
so that no transformation occurs, the VW ultimately moving by 360~nm.
Contrarily to automotion by transformation where the displacement
is quantized (strictly if no vortex core switching occurs), this pure 
STT displacement depends on $J$, $\beta$ and pulse length.
An experimental comparison of the displacements under DW structure
transformation and under pure STT is presented in 
Fig.~\ref{fig:stt}.
For the short pulse used here (1~ns), the displacement
due to automotion by transformation clearly dominates.
Note also that the value of the VW displacement is consistent with
the value $\beta= 0.08$ chosen for the simulations.

Thus, we have experimentally demonstrated that DW automotion during
a structure transformation gives rise, in real samples, to 
measurable displacements.
A signature of the effect is the direct relation with the change of
generalized wall angle, as checked by high resolution
MFM imaging.
The situation chosen for this demonstration was that of a metastable 
ATW, frequently encountered in experiments, but the conclusions are 
general: a DW transformation, whatever its cause (field \cite{field}, 
current, etc) that modifies the generalized wall 
angle $\Phi$, leads to an intrinsic and large (for small damping) 
DW displacement.
Note also that the same physical arguments should apply 
to samples with perpendicular magnetization.
We thus propose that, once the damping constant relevant for DW dynamics 
is known, the magnitude of the displacement by automotion be taken as 
a measure of sample quality.
When a DW transformation occurs, automotion dominates as soon as 
the duration of excitation is short compared to the relaxation time 
of the DW structure: the linear momentum bestowed to the DW when
it starts transforming is very large as $\Phi$ is far from
equilibrium, and it is damped over the relaxation time.
In particular, the resulting displacement is of a different nature,
and much larger than that due to the pulse risetime effect 
\cite{Bocklage09}, derived in the small $\Phi$ approximation that
forgets the possibility of a transformation: for the short pulses 
(compared to the relaxation time of the DW structure) applied here,
the force due to the pulse risetime effect is a $\delta'$ that causes 
no motion.

Automotion may impact the measurements of DW displacement under 
pulsed excitation (field, or spin-polarized current).
This is especially true for experiments in which a new DW is created 
prior to the application of each pulse: if this DW is
not in the stable state, a transformation with a large $\Phi$
angle change will occur, resulting in a DW displacement as large
as what has been measured here.
This explains the apparent "better mobility" under STT of a
transformed DW.
Thus, the absence of DW structure transformation has to be
checked by imaging, but this is not always possible.
Alternatively, automotion by transformation can be seen as a
non-zero displacement at zero extrapolated excitation (varying duration
or, as shown here in Fig.~\ref{fig:gene}, amplitude).
It implies that apparent velocities may be meaningless for
quantitatively evaluating the spin transfer torque terms.

This work was supported by the ANR-07-NANO-034 "Dynawall" project.
We thank M. Aprili, J. Gabelli and T. Devolder for their assistance.

\bibliography{auto7cm3}

\begin{thebibliography}{31}
\expandafter\ifx\csname natexlab\endcsname\relax\def\natexlab#1{#1}\fi
\expandafter\ifx\csname bibnamefont\endcsname\relax
  \def\bibnamefont#1{#1}\fi
\expandafter\ifx\csname bibfnamefont\endcsname\relax
  \def\bibfnamefont#1{#1}\fi
\expandafter\ifx\csname citenamefont\endcsname\relax
  \def\citenamefont#1{#1}\fi
\expandafter\ifx\csname url\endcsname\relax
  \def\url#1{\texttt{#1}}\fi
\expandafter\ifx\csname urlprefix\endcsname\relax\def\urlprefix{URL }\fi
\providecommand{\bibinfo}[2]{#2}
\providecommand{\eprint}[2][]{\url{#2}}

\bibitem[{\citenamefont{Berger}(1984)}]{Berger84}
\bibinfo{author}{\bibfnamefont{L.}~\bibnamefont{Berger}}, \bibinfo{journal}{J.\
  Appl.\ Phys.} \textbf{\bibinfo{volume}{55}}, \bibinfo{pages}{1954}
  (\bibinfo{year}{1984}).

\bibitem[{\citenamefont{Parkin et~al.}(2008)\citenamefont{Parkin, Hayashi, and
  Thomas}}]{Parkin08}
\bibinfo{author}{\bibfnamefont{S.}~\bibnamefont{Parkin}},
  \bibinfo{author}{\bibfnamefont{M.}~\bibnamefont{Hayashi}}, \bibnamefont{and}
  \bibinfo{author}{\bibfnamefont{L.}~\bibnamefont{Thomas}},
  \bibinfo{journal}{Science} \textbf{\bibinfo{volume}{320}},
  \bibinfo{pages}{190} (\bibinfo{year}{2008}).

\bibitem[{\citenamefont{Tserkovnyak et~al.}(2008)\citenamefont{Tserkovnyak,
  Brataas, and Bauer}}]{Tserkovnyak08}
\bibinfo{author}{\bibfnamefont{Y.}~\bibnamefont{Tserkovnyak}},
  \bibinfo{author}{\bibfnamefont{A.}~\bibnamefont{Brataas}}, \bibnamefont{and}
  \bibinfo{author}{\bibfnamefont{G.}~\bibnamefont{Bauer}},
  \bibinfo{journal}{J.\ Magn.\ Magn.\ Mater.} \textbf{\bibinfo{volume}{320}},
  \bibinfo{pages}{1282} (\bibinfo{year}{2008}).

\bibitem[{\citenamefont{Yamaguchi et~al.}(2004)\citenamefont{Yamaguchi, Ono,
  Nasu, Miyake, Mibu, and Shinjo}}]{Yamaguchi04}
\bibinfo{author}{\bibfnamefont{A.}~\bibnamefont{Yamaguchi}},
  \bibinfo{author}{\bibfnamefont{T.}~\bibnamefont{Ono}},
  \bibinfo{author}{\bibfnamefont{S.}~\bibnamefont{Nasu}},
  \bibinfo{author}{\bibfnamefont{K.}~\bibnamefont{Miyake}},
  \bibinfo{author}{\bibfnamefont{K.}~\bibnamefont{Mibu}}, \bibnamefont{and}
  \bibinfo{author}{\bibfnamefont{T.}~\bibnamefont{Shinjo}},
  \bibinfo{journal}{Phys.\ Rev.\ Lett.} \textbf{\bibinfo{volume}{92}},
  \bibinfo{pages}{077205} (\bibinfo{year}{2004}).

\bibitem[{\citenamefont{Kl{{\"a}}ui et~al.}(2005)\citenamefont{Kl{{\"a}}ui,
  Jubert, Allenspach, Bischof, Bland, Faini, R{{\"u}}diger, Vaz, Vila, and
  Vouille}}]{Klaui05}
\bibinfo{author}{\bibfnamefont{M.}~\bibnamefont{Kl{{\"a}}ui}},
  \bibinfo{author}{\bibfnamefont{P.}~\bibnamefont{Jubert}},
  \bibinfo{author}{\bibfnamefont{R.}~\bibnamefont{Allenspach}},
  \bibinfo{author}{\bibfnamefont{A.}~\bibnamefont{Bischof}},
  \bibinfo{author}{\bibfnamefont{J.}~\bibnamefont{Bland}},
  \bibinfo{author}{\bibfnamefont{G.}~\bibnamefont{Faini}},
  \bibinfo{author}{\bibfnamefont{U.}~\bibnamefont{R{{\"u}}diger}},
  \bibinfo{author}{\bibfnamefont{C.}~\bibnamefont{Vaz}},
  \bibinfo{author}{\bibfnamefont{L.}~\bibnamefont{Vila}}, \bibnamefont{and}
  \bibinfo{author}{\bibfnamefont{C.}~\bibnamefont{Vouille}},
  \bibinfo{journal}{Phys.\ Rev.\ Lett.} \textbf{\bibinfo{volume}{95}},
  \bibinfo{pages}{026601} (\bibinfo{year}{2005}).

\bibitem[{\citenamefont{Meier et~al.}(2007)\citenamefont{Meier, Bolte, Eiselt,
  Kr{\"{u}}ger, Kim, and Fischer}}]{Meier07}
\bibinfo{author}{\bibfnamefont{G.}~\bibnamefont{Meier}},
  \bibinfo{author}{\bibfnamefont{M.}~\bibnamefont{Bolte}},
  \bibinfo{author}{\bibfnamefont{R.}~\bibnamefont{Eiselt}},
  \bibinfo{author}{\bibfnamefont{B.}~\bibnamefont{Kr{\"{u}}ger}},
  \bibinfo{author}{\bibfnamefont{D.-H.} \bibnamefont{Kim}}, \bibnamefont{and}
  \bibinfo{author}{\bibfnamefont{P.}~\bibnamefont{Fischer}},
  \bibinfo{journal}{Phys.\ Rev.\ Lett.} \textbf{\bibinfo{volume}{98}},
  \bibinfo{pages}{187202} (\bibinfo{year}{2007}).

\bibitem[{\citenamefont{Vlaminck and Bailleul}(2008)}]{Vlaminck08}
\bibinfo{author}{\bibfnamefont{V.}~\bibnamefont{Vlaminck}} \bibnamefont{and}
  \bibinfo{author}{\bibfnamefont{M.}~\bibnamefont{Bailleul}},
  \bibinfo{journal}{Science} \textbf{\bibinfo{volume}{322}},
  \bibinfo{pages}{410} (\bibinfo{year}{2008}).

\bibitem[{\citenamefont{Bazaliy et~al.}(1998)\citenamefont{Bazaliy, Jones, and
  Zhang}}]{Bazaliy98}
\bibinfo{author}{\bibfnamefont{Y.~B.} \bibnamefont{Bazaliy}},
  \bibinfo{author}{\bibfnamefont{B.}~\bibnamefont{Jones}}, \bibnamefont{and}
  \bibinfo{author}{\bibfnamefont{S.-C.} \bibnamefont{Zhang}},
  \bibinfo{journal}{Phys.\ Rev.\ B} \textbf{\bibinfo{volume}{57}},
  \bibinfo{pages}{R3213} (\bibinfo{year}{1998}).

\bibitem[{\citenamefont{Zhang and Li}(2004)}]{Zhang04}
\bibinfo{author}{\bibfnamefont{S.}~\bibnamefont{Zhang}} \bibnamefont{and}
  \bibinfo{author}{\bibfnamefont{Z.}~\bibnamefont{Li}},
  \bibinfo{journal}{Phys.\ Rev.\ Lett.} \textbf{\bibinfo{volume}{93}},
  \bibinfo{pages}{127204} (\bibinfo{year}{2004}).

\bibitem[{\citenamefont{Tatara and Kohno}(2004)}]{Tatara04}
\bibinfo{author}{\bibfnamefont{G.}~\bibnamefont{Tatara}} \bibnamefont{and}
  \bibinfo{author}{\bibfnamefont{H.}~\bibnamefont{Kohno}},
  \bibinfo{journal}{Phys.\ Rev.\ Lett.} \textbf{\bibinfo{volume}{92}},
  \bibinfo{pages}{086601} (\bibinfo{year}{2004}).

\bibitem[{\citenamefont{Thiaville et~al.}(2005)\citenamefont{Thiaville,
  Nakatani, Miltat, and Suzuki}}]{Thiaville05}
\bibinfo{author}{\bibfnamefont{A.}~\bibnamefont{Thiaville}},
  \bibinfo{author}{\bibfnamefont{Y.}~\bibnamefont{Nakatani}},
  \bibinfo{author}{\bibfnamefont{J.}~\bibnamefont{Miltat}}, \bibnamefont{and}
  \bibinfo{author}{\bibfnamefont{Y.}~\bibnamefont{Suzuki}},
  \bibinfo{journal}{Europhys.\ Lett.} \textbf{\bibinfo{volume}{69}},
  \bibinfo{pages}{990} (\bibinfo{year}{2005}).

\bibitem[{\citenamefont{Jubert et~al.}(2006)\citenamefont{Jubert, Kl{\"{a}}ui,
  Bischof, R{\"{u}}diger, and Allenspach}}]{Jubert06}
\bibinfo{author}{\bibfnamefont{P.-O.} \bibnamefont{Jubert}},
  \bibinfo{author}{\bibfnamefont{M.}~\bibnamefont{Kl{\"{a}}ui}},
  \bibinfo{author}{\bibfnamefont{A.}~\bibnamefont{Bischof}},
  \bibinfo{author}{\bibfnamefont{U.}~\bibnamefont{R{\"{u}}diger}},
  \bibnamefont{and}
  \bibinfo{author}{\bibfnamefont{R.}~\bibnamefont{Allenspach}},
  \bibinfo{journal}{J.\ Appl.\ Phys.} \textbf{\bibinfo{volume}{99}},
  \bibinfo{pages}{08G523} (\bibinfo{year}{2006}).

\bibitem[{\citenamefont{Yang and Erskine}(2007)}]{Yang07}
\bibinfo{author}{\bibfnamefont{S.}~\bibnamefont{Yang}} \bibnamefont{and}
  \bibinfo{author}{\bibfnamefont{J.}~\bibnamefont{Erskine}},
  \bibinfo{journal}{Phys.\ Rev.\ B} \textbf{\bibinfo{volume}{75}},
  \bibinfo{pages}{220403(R)} (\bibinfo{year}{2007}).

\bibitem[{\citenamefont{Hayashi et~al.}(2007)\citenamefont{Hayashi, Thomas,
  Rettner, Moriya, Bazaliy, and Parkin}}]{Hayashi07}
\bibinfo{author}{\bibfnamefont{M.}~\bibnamefont{Hayashi}},
  \bibinfo{author}{\bibfnamefont{L.}~\bibnamefont{Thomas}},
  \bibinfo{author}{\bibfnamefont{C.}~\bibnamefont{Rettner}},
  \bibinfo{author}{\bibfnamefont{R.}~\bibnamefont{Moriya}},
  \bibinfo{author}{\bibfnamefont{Y.}~\bibnamefont{Bazaliy}}, \bibnamefont{and}
  \bibinfo{author}{\bibfnamefont{S.}~\bibnamefont{Parkin}},
  \bibinfo{journal}{Phys.\ Rev.\ Lett.} \textbf{\bibinfo{volume}{98}},
  \bibinfo{pages}{037204} (\bibinfo{year}{2007}).

\bibitem[{\citenamefont{Heyne et~al.}(2010)\citenamefont{Heyne, Rhensius,
  Bisig, Krzyk, Punke, Kl{\"{a}}ui, Heyderman, {Le Guyader}, and
  Nolting}}]{Heyne10}
\bibinfo{author}{\bibfnamefont{L.}~\bibnamefont{Heyne}},
  \bibinfo{author}{\bibfnamefont{J.}~\bibnamefont{Rhensius}},
  \bibinfo{author}{\bibfnamefont{A.}~\bibnamefont{Bisig}},
  \bibinfo{author}{\bibfnamefont{S.}~\bibnamefont{Krzyk}},
  \bibinfo{author}{\bibfnamefont{P.}~\bibnamefont{Punke}},
  \bibinfo{author}{\bibfnamefont{M.}~\bibnamefont{Kl{\"{a}}ui}},
  \bibinfo{author}{\bibfnamefont{L.}~\bibnamefont{Heyderman}},
  \bibinfo{author}{\bibfnamefont{L.}~\bibnamefont{{Le Guyader}}},
  \bibnamefont{and} \bibinfo{author}{\bibfnamefont{F.}~\bibnamefont{Nolting}},
  \bibinfo{journal}{Appl.\ Phys.\ Lett.} \textbf{\bibinfo{volume}{96}},
  \bibinfo{pages}{032504} (\bibinfo{year}{2010}).

\bibitem[{\citenamefont{Bocklage et~al.}(2009)\citenamefont{Bocklage,
  Kr{\"{u}}ger, Matsuyama, Bolte, Merkt, Pfannkuche, and Meier}}]{Bocklage09}
\bibinfo{author}{\bibfnamefont{L.}~\bibnamefont{Bocklage}},
  \bibinfo{author}{\bibfnamefont{B.}~\bibnamefont{Kr{\"{u}}ger}},
  \bibinfo{author}{\bibfnamefont{T.}~\bibnamefont{Matsuyama}},
  \bibinfo{author}{\bibfnamefont{M.}~\bibnamefont{Bolte}},
  \bibinfo{author}{\bibfnamefont{U.}~\bibnamefont{Merkt}},
  \bibinfo{author}{\bibfnamefont{D.}~\bibnamefont{Pfannkuche}},
  \bibnamefont{and} \bibinfo{author}{\bibfnamefont{G.}~\bibnamefont{Meier}},
  \bibinfo{journal}{Phys.\ Rev.\ Lett.} \textbf{\bibinfo{volume}{103}},
  \bibinfo{pages}{197204} (\bibinfo{year}{2009}).

\bibitem[{\citenamefont{Li and Zhang}(2004)}]{Li04}
\bibinfo{author}{\bibfnamefont{Z.}~\bibnamefont{Li}} \bibnamefont{and}
  \bibinfo{author}{\bibfnamefont{S.}~\bibnamefont{Zhang}},
  \bibinfo{journal}{Phys.\ Rev.\ Lett.} \textbf{\bibinfo{volume}{92}},
  \bibinfo{pages}{207203} (\bibinfo{year}{2004}).

\bibitem[{\citenamefont{Thiaville et~al.}(2007)\citenamefont{Thiaville,
  Nakatani, Pi{\'{e}}chon, Miltat, and Ono}}]{Thiaville07}
\bibinfo{author}{\bibfnamefont{A.}~\bibnamefont{Thiaville}},
  \bibinfo{author}{\bibfnamefont{Y.}~\bibnamefont{Nakatani}},
  \bibinfo{author}{\bibfnamefont{F.}~\bibnamefont{Pi{\'{e}}chon}},
  \bibinfo{author}{\bibfnamefont{J.}~\bibnamefont{Miltat}}, \bibnamefont{and}
  \bibinfo{author}{\bibfnamefont{T.}~\bibnamefont{Ono}},
  \bibinfo{journal}{Eur.\ Phys.\ J. B} \textbf{\bibinfo{volume}{60}},
  \bibinfo{pages}{15} (\bibinfo{year}{2007}).

\bibitem[{\citenamefont{Stein and Feldtkeller}(1967)}]{Stein67}
\bibinfo{author}{\bibfnamefont{K.}~\bibnamefont{Stein}} \bibnamefont{and}
  \bibinfo{author}{\bibfnamefont{E.}~\bibnamefont{Feldtkeller}},
  \bibinfo{journal}{J.\ Appl.\ Phys.} \textbf{\bibinfo{volume}{38}},
  \bibinfo{pages}{4401} (\bibinfo{year}{1967}).

\bibitem[{\citenamefont{Malozemoff and Slonczewski}(1979)}]{Malozemoff79}
\bibinfo{author}{\bibfnamefont{A.}~\bibnamefont{Malozemoff}} \bibnamefont{and}
  \bibinfo{author}{\bibfnamefont{J.}~\bibnamefont{Slonczewski}},
  \emph{\bibinfo{title}{Magnetic Domain Walls in Bubble Materials}}
  (\bibinfo{publisher}{Academic Press}, \bibinfo{address}{New York},
  \bibinfo{year}{1979}).

\bibitem[{\citenamefont{Thiele}(1976)}]{Thiele76}
\bibinfo{author}{\bibfnamefont{A.}~\bibnamefont{Thiele}}, \bibinfo{journal}{J.\
  Appl.\ Phys.} \textbf{\bibinfo{volume}{47}}, \bibinfo{pages}{2759}
  (\bibinfo{year}{1976}).

\bibitem[{\citenamefont{Slonczewski}(1979)}]{Slonczewski79}
\bibinfo{author}{\bibfnamefont{J.}~\bibnamefont{Slonczewski}},
  \bibinfo{journal}{J.\ Magn.\ Magn.\ Mater.} \textbf{\bibinfo{volume}{12}},
  \bibinfo{pages}{108} (\bibinfo{year}{1979}).

\bibitem[{\citenamefont{Heyne et~al.}(2008)\citenamefont{Heyne, Kl{{\"a}}ui,
  Backes, M{\"{o}}hrke, Moore, Kimling, Boulle, R{{\"u}}diger, Heyderman,
  Rodr{{\'i}}guez et~al.}}]{Heyne08b}
\bibinfo{author}{\bibfnamefont{L.}~\bibnamefont{Heyne}},
  \bibinfo{author}{\bibfnamefont{M.}~\bibnamefont{Kl{{\"a}}ui}},
  \bibinfo{author}{\bibfnamefont{D.}~\bibnamefont{Backes}},
  \bibinfo{author}{\bibfnamefont{P.}~\bibnamefont{M{\"{o}}hrke}},
  \bibinfo{author}{\bibfnamefont{T.}~\bibnamefont{Moore}},
  \bibinfo{author}{\bibfnamefont{J.}~\bibnamefont{Kimling}},
  \bibinfo{author}{\bibfnamefont{O.}~\bibnamefont{Boulle}},
  \bibinfo{author}{\bibfnamefont{U.}~\bibnamefont{R{{\"u}}diger}},
  \bibinfo{author}{\bibfnamefont{L.}~\bibnamefont{Heyderman}},
  \bibinfo{author}{\bibfnamefont{A.~F.} \bibnamefont{Rodr{{\'i}}guez}},
  \bibnamefont{et~al.}, \bibinfo{journal}{J.\ Appl.\ Phys.}
  \textbf{\bibinfo{volume}{103}}, \bibinfo{pages}{07D928}
  (\bibinfo{year}{2008}).

\bibitem[{\citenamefont{Nakatani et~al.}(2005)\citenamefont{Nakatani,
  Thiaville, and Miltat}}]{Nakatani05}
\bibinfo{author}{\bibfnamefont{Y.}~\bibnamefont{Nakatani}},
  \bibinfo{author}{\bibfnamefont{A.}~\bibnamefont{Thiaville}},
  \bibnamefont{and} \bibinfo{author}{\bibfnamefont{J.}~\bibnamefont{Miltat}},
  \bibinfo{journal}{J.\ Magn.\ Magn.\ Mater.}
  \textbf{\bibinfo{volume}{290-291}}, \bibinfo{pages}{750}
  (\bibinfo{year}{2005}).

\bibitem[{\citenamefont{Kl{{\"a}}ui et~al.}(2004)\citenamefont{Kl{{\"a}}ui,
  Vaz, Bland, Heyderman, Nolting, Pavlovska, Bauer, Cherifi, Heun, and
  Locatelli}}]{Klaui04}
\bibinfo{author}{\bibfnamefont{M.}~\bibnamefont{Kl{{\"a}}ui}},
  \bibinfo{author}{\bibfnamefont{C.}~\bibnamefont{Vaz}},
  \bibinfo{author}{\bibfnamefont{J.}~\bibnamefont{Bland}},
  \bibinfo{author}{\bibfnamefont{L.}~\bibnamefont{Heyderman}},
  \bibinfo{author}{\bibfnamefont{F.}~\bibnamefont{Nolting}},
  \bibinfo{author}{\bibfnamefont{A.}~\bibnamefont{Pavlovska}},
  \bibinfo{author}{\bibfnamefont{E.}~\bibnamefont{Bauer}},
  \bibinfo{author}{\bibfnamefont{S.}~\bibnamefont{Cherifi}},
  \bibinfo{author}{\bibfnamefont{S.}~\bibnamefont{Heun}}, \bibnamefont{and}
  \bibinfo{author}{\bibfnamefont{A.}~\bibnamefont{Locatelli}},
  \bibinfo{journal}{Appl.\ Phys.\ Lett.} \textbf{\bibinfo{volume}{85}},
  \bibinfo{pages}{5637} (\bibinfo{year}{2004}).

\bibitem[{\citenamefont{Thiaville and Nakatani}(2009)}]{Thiaville09}
\bibinfo{author}{\bibfnamefont{A.}~\bibnamefont{Thiaville}} \bibnamefont{and}
  \bibinfo{author}{\bibfnamefont{Y.}~\bibnamefont{Nakatani}},
  \emph{\bibinfo{title}{Nanomagnetism and Spintronics}}
  (\bibinfo{publisher}{Elsevier}, \bibinfo{address}{Amsterdam},
  \bibinfo{year}{2009}), chap.~\bibinfo{chapter}{6}, pp.
  \bibinfo{pages}{231--276}.

\bibitem[{\citenamefont{Miltat and Donahue}(2007)}]{Miltat07}
\bibinfo{author}{\bibfnamefont{J.}~\bibnamefont{Miltat}} \bibnamefont{and}
  \bibinfo{author}{\bibfnamefont{M.}~\bibnamefont{Donahue}},
  \emph{\bibinfo{title}{Handbook of Magnetism and Advanced Magnetic Materials}}
  (\bibinfo{publisher}{Wiley}, \bibinfo{address}{New York},
  \bibinfo{year}{2007}), vol.~\bibinfo{volume}{2}, pp.
  \bibinfo{pages}{742--764}.

\bibitem[{\citenamefont{Nakatani et~al.}(2003)\citenamefont{Nakatani,
  Thiaville, and Miltat}}]{Nakatani03}
\bibinfo{author}{\bibfnamefont{Y.}~\bibnamefont{Nakatani}},
  \bibinfo{author}{\bibfnamefont{A.}~\bibnamefont{Thiaville}},
  \bibnamefont{and} \bibinfo{author}{\bibfnamefont{J.}~\bibnamefont{Miltat}},
  \bibinfo{journal}{Nature Mater.} \textbf{\bibinfo{volume}{2}},
  \bibinfo{pages}{521} (\bibinfo{year}{2003}).

\bibitem[{\citenamefont{Min et~al.}(2010)\citenamefont{Min, McMichael, Donahue,
  Miltat, and Stiles}}]{Min10}
\bibinfo{author}{\bibfnamefont{H.}~\bibnamefont{Min}},
  \bibinfo{author}{\bibfnamefont{R.}~\bibnamefont{McMichael}},
  \bibinfo{author}{\bibfnamefont{M.}~\bibnamefont{Donahue}},
  \bibinfo{author}{\bibfnamefont{J.}~\bibnamefont{Miltat}}, \bibnamefont{and}
  \bibinfo{author}{\bibfnamefont{M.}~\bibnamefont{Stiles}},
  \bibinfo{journal}{Phys.\ Rev.\ Lett.} \textbf{\bibinfo{volume}{104}},
  \bibinfo{pages}{217201} (\bibinfo{year}{2010}).

\bibitem[{onl()}]{online-movies}
\bibinfo{note}{See the related EPAPS document for movies created from the
  micromagnetic simulations.}

\bibitem[{fie()}]{field}
\bibinfo{note}{It was checked that the same displacements were measured after
  submitting the metastable ATW to nanosecond transverse field pulses, using a
  nanostrip coated by a thick conductive layer.}

\end{thebibliography}

\end{document}